\pdfoutput=1
\documentclass[a4 paper, 10pt,twoside]{article}
\usepackage{xcolor}
\usepackage{graphicx}
\usepackage{latexsym}
\usepackage{url}
\usepackage{xcolor}

\def\2F1{_2F_1}
\def\h0units{\mathrm{km\,s^{-1}\,Mpc^{-1}}}

\newcommand{\om}{\Omega_{\rm M}}
\newcommand{\ok}{\Omega_K}
\newcommand{\ola}{\Omega_{\Lambda}}
\newcommand{\ode}{\Omega_{DE}}
\newcommand{\dl}{d_{\rm{L}}}
\newcommand{\dlsette}{d_{\rm{L,7}}}
\newcommand{\dlnum}{d_{\rm{L,num}}}

\begin{document}
\def\aap{A\&A\,  }
\def \aapr{The Astronomy and Astrophysics Review}
\def\aaps{A\&AS  }
\def\acp{Anal. Cell. Pathol. } 
\def\aj{AJ  }
\def\aplett{Astrophys. Lett.\,  }
\def\apj{ApJ\,  }
\def\apjl{ApJ Letters,  }
\def\apjs{ApJS  }
\def\apss{Astrophysics and Space Science  }
\def\araa{ARA\&A  }
\def\azh{AZh}
\def\bain{BAN  }
\def\baas{Bulletin of the American Astronomical Society}
\def\cjaa{Chinese Astronomy and Astrophysics}
\def\eup{Europhys. Lett.  }
\def\fcp{Fundamentals of Cosmic Physics}
\def\iaucirc{IAU circ.  } 
\def\icarus{Icarus} 
\def\jaa{J. Astrophys. Astr.  }
\def\jpc{J. Phys. C  } 
\def\JPG{J. Phys. G\,  }
\def\jsp{J. Stat. Phys  } 
\def\jcap{Journal of Cosmology and Astroparticle Physic  } 
\def\jcp{J. Comput. Phys.  } 
\def\jcpp{Journal of Chemical  Physics  } 
\def\jrasc{JRASC  } 
\def\mnras{MNRAS\,  }

\def\na {New Astronomy\,  }
\def\nat{Nature\,  }
\def\npb{Nuc. Phys. B   }
\def\oe{Optic  Express }
\def\pasj{PASJ\,  }
\def\solphys{Sol. Phys.\,  }
\def\planss{Planet. Space Sci.  }
\def\pasp{PASP  }
\def\pasa{PASA  }
\def\POF{Physics of Fluids  }
\def\physrep{Phys. Rep.\,  }
\def\pla{Phys. Lett. A   }
\def\pra{Phys. Rev. A   }
\def\prb{Phys. Rev. B   }
\def\prc{Phys. Rev. C   }
\def\prd{Phys. Rev. D   }
\def\pre{Phys. Rev. E   }
\def\prl{Phys. Rev. Lett.    }
\def\physa{Phys. A    }
\def\rmp{Rev. Mod. Phys.  }
\def\rmxaa{Revista Mexicana de Astronomia y Astrofisica} 
\def\skytel{Sky and Telescope}
\def\ssr{Space Science Reviews} 
\def\rpp{Rep.Prog.Phys.   }
\def\sovast{Soviet Astronomy} 
\def\za{Z. Astrophys.  } 
\def\zap{Zeitschrift fur Astrophysik} 
\title
{
The distance modulus  in 
dark energy and Cardassian cosmologies via  the
hypergeometric function
}

\vspace{2pc}
\author     {Lorenzo  Zaninetti  \\
Physics Department,
 via P.Giuria 1,\\ I-10125 Turin,Italy\\ 
Email: zaninetti@ph.unito.it}
\maketitle
\begin{abstract}
The presence  of the dark energy allows both the acceleration 
and the expansion of the universe.
In the case of 
a  constant equation of state for dark energy
we derived   
an analytical solution  for the Hubble radius  
in terms of the hypergeometric function.
An approximate Taylor expansion 
of order seven
is derived for both 
the constant 
and  
the variable equation of state for dark energy.
In the case of the Cardassian cosmology we also derived
an analytical solution for the Hubble radius 
in terms of the hypergeometric function.
The astronomical  samples  of the 
distance  modulus  
for Supernova (SN)  of type Ia
allows the derivation of the involved cosmological
in the case of 
constant equation 
of state,
variable   equation 
of state
and 
Cardassian cosmology.
\end{abstract}

\vspace{2pc}
\vspace{2pc}

\maketitle

\section{Introduction}
The name dark energy  started to be used 
by \cite{Turner1999} in order to explain both the expansion
and both the  acceleration of the universe.
In a few years the dark energy  was  widely used as   
a  cosmological model to be tested.
Many review papers has been written;
we select among others  
a general review by  
\cite{Huterer2018}
and  a theoretical review by   \cite{Brax2018}.
The term wCDM has been introduced to classify  
the case of constant equation of state 
and we will use in the following 
wzCDM to classify the variable equation of state.
The Cardassian cosmology started with \cite{Freese2002}
and was introduced in order to model both the expansion 
and the acceleration of the universe, the name from  
a humanoid race in Star Trek.
As an example  \cite{Magana2018} derived the cosmological parameters 
for 
the original Cardassian expansion and 
the modified polytropic Cardassian expansion.
The cosmological theories can be tested
on the samples of  Supernova (SN)  of type Ia.
The first sample to be used to derive the cosmological parameters
contained 
7   SNs, see \cite{Perlmutter1997},
the second one contained 34 SNs,
see Figure 4 in \cite{Riess1998}
and the third  one
contained 42  SNs, see  \cite{Perlmutter1999}.
The above historical samples 
allowed to derive the cosmological parameters
for  the expanding and accelerating universe.
At the moment of writing 
the astronomical research is focused on 
value of the distance modulus versus the redshift:
the Union 2.1 compilation contains 580 SNs, 
see \cite{Suzuki2012},
and    
the joint light-curve analysis (JLA)
contains  740 SNs, see \cite{Betoule2014}.
The above  observations can be done up to a limited 
value in redshift $z \approx 1.7$,
we therefore speak of evaluation of the distance modulus
at low redshift.
This limited range can be extended up 
$z \approx 8 $, the high redshift region,  
analyzing  the
Gamma-Ray Burst (GRB) and, as an example,   
 \cite {Wei2010} has 
derived the distance modulus for 59 
calibrated high-redshift GRBs, the  
so called  the "Hymnium" GRBs sample.
This paper 
reviews 
in Section \ref{review} the $\Lambda$CDM cosmology,
evaluates 
the basic integral of wCDM cosmology in Section \ref{wcdm},
introduces a Taylor  expansion for the 
basic integral of wzCDM cosmology in Section \ref{wzcdm} and
analyzes the Cardassian model  in Section \ref{seccardassian}.
The  parameters which characterizes the three cosmologies 
 are 
derived via the Levenberg--Marquardt  method 
in Section \ref{dm}.

\section{Preliminaries}

This section reviews the 
$\Lambda$CDM cosmology and the 
adopted statistics.

\subsection{The standard cosmology}
\label{review}

In $\Lambda$CDM cosmology
the {\em Hubble
distance\/} $D_{\rm H}$
is defined as
\begin{equation}
\label{eq:dh}
D_{\rm H}\equiv\frac{c}{H_0}
\quad .
\end{equation}
The first parameter is 
 $\om$
\begin{equation}
\om = \frac{8\pi\,G\,\rho_0}{3\,H_0^2}
\quad ,
\end{equation}
where $G$ is the Newtonian gravitational constant,
$H_0$ is the Hubble constant 
 and
$\rho_0$ is the mass density at the present time.
The second parameter is $\ola$
\begin{equation}
\ola\equiv\frac{\Lambda\,c^2}{3\,H_0^2}
\quad ,
\end{equation}
where $\Lambda$ is the cosmological constant,
see \cite{Peebles1993}.
These two parameters are connected with the
curvature $\ok$ by
\begin{equation}
\om+\ola+\ok= 1
\quad .
\end{equation}
The  comoving distance, $D_{\rm C}$,  is
\begin{equation}
D_{\rm C} = D_{\rm H}\,\int_0^z\frac{dz'}{E(z')}
\label{integralezstandard}
\end{equation}
where $E(z)$ is the "Hubble function"
\begin{equation}
\label{eq:ez}
E(z) = \sqrt{\om\,(1+z)^3+\ok\,(1+z)^2+\ola}
\quad .
\end{equation}
In the case of $\ok=$ we have the flat case.

\subsection{The statistics}

The adopted  statistical parameters  are 
the percent error, $\delta$, between
theoretical value and approximated value,
the merit function $\chi^2$ evaluated
as
\begin{equation} 
\chi^2 =\sum_{i=1}^N \Big 
[ \frac{ y_{i,theo}-y_{i,obs} }{ \sigma_i } \Big ]^2
\label{chiquaredef}
\end{equation}
where $y_{i,obs}$ and $\sigma_i$   represent
the observed value and its error at position $i$
and $y_{i,theo}$ the theoretical value at position $i$,
the  reduced  merit function $\chi_{red}^2$,
the Akaike information criterion
(AIC),
the number of degrees  of freedom
$NF=n-k$ where 
$n$ is the number of bins and $k$ is the number of parameters and
the goodness  of the fit  as expressed by
the probability $Q$.

\section{Constant equation of state}

\label{wcdm}
In dark matter cosmology, wCDM,  
the Hubble radius   is    
\begin{equation}
d_H(z;\om,w,\ode) =
 \frac {1}{
\sqrt { 
\left( 1+z \right) ^{3}{\it \om}+{\it \ode}\, \left( 1+z
 \right) ^{3+3\,w}
}
 }
\quad ,
\end{equation}
where  $w$ parametrizes  the dark energy and is constant,
see equation(3.4) in  \cite{Tripathi2017} or
equation(18) in \cite{Wei2015} for the luminosity distance.

In flat cosmology
\begin{equation}
\om +\ode=1
\quad ,
\end{equation}
and the  Hubble radius becomes
\begin{equation}
d_H(z;\om,w) =
 \frac {1}
{
\sqrt { \left( 1+z \right) ^{3}{\it \om}+ \left( 1-{\it \om} \right) 
 \left( 1+z \right) ^{3+3\,w}}
 }
\quad .
\end{equation}
The indefinite integral  in the  variable  $z$ of the 
above  Hubble radius, $Iz$,  is 
\begin{equation}
Iz(z;\om,w) =
\int d_H(z;\om,w)  dz 
\quad .
\label{integralez}
\end{equation}

\subsection{The analytical solution }

In order to solve the indefinite integral we  perform a change of variable
$1+z=t^{1/3}$
\begin{equation}
Iz(t;\om,w) =
\frac{1}{3} \int \!{\frac {1}{\sqrt {-t \left(  \left( -1+{\it \om} \right)
{t}^{w}
-{\it \om} \right) }{t}^{2/3}}}\,{\rm d}t
\quad .
\label{integralet}
\end{equation}
The indefinite integral is
\begin{equation}
Iz(t;\om,w) =
\frac
{
-2\,
{\mbox{$_2$F$_1$}(\frac{1}{2},-\frac{1}{6}\,{w}^{-1};\,1-\frac{1}{6}\,{w}^{-1};\,-{\frac {{t}^{w}
- \left( 1-{\it \om} \right) }{{\it \om}}})}
}
{
\sqrt {{\it \om}}\sqrt [6]{t}
}
\quad ,
\label{hubbleintegral}
\end{equation}
where ${\2F1(a,b;\,c;\,z)}$ is the
regularized hypergeometric
function, see Appendix \ref{appendixb}.
This dependence of the above integral upon  the hypergeometric function has  been
recognized but not developed by \cite{Wickramasinghe2010}.
 
We now return to the  variable $z$, the redshift,
and the indefinite integral becomes
\begin{eqnarray}
Iz(z;\om,w) = \nonumber \\
\frac
{
-2\,
{\mbox{$_2$F$_1$}(\frac{1}{2},-\frac{1}{6}\,{w}^{-1};\,1-\frac{1}{6}\,{w}^{-1};\,-{\frac { \left(
- {z}^{3}+3\,{z}^{2}+3\,z+1 \right) ^{w} \left( 1-{\it \om} \right) }{{\it
- \om}}})}
}
{
\sqrt {{\it \om}}\sqrt [6]{{z}^{3}+3\,{z}^{2}+3\,z+1}
}
\quad .
\end{eqnarray}
We denote by $F(z;\om,w)$ the definite integral
\begin{equation}
F(z;\om,w)  = Iz(z=z;\om,w)  - Iz(z=0;\om,w)
\quad .
\label{definitefz}
\end{equation} 

\subsection{The Taylor expansion}

\begin{figure}
\includegraphics[width=8cm]{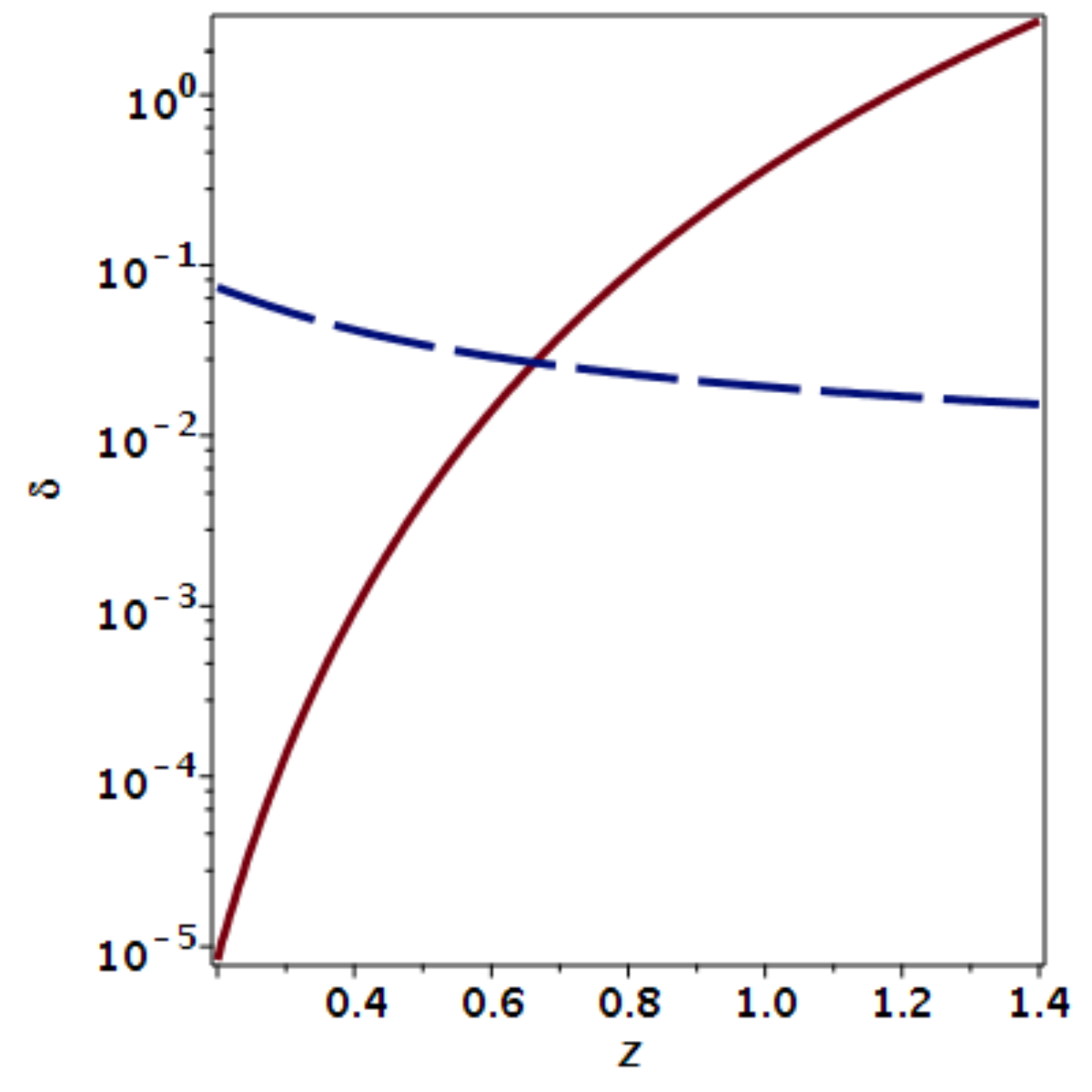}
\caption
{
Numerical values of 
$\delta_I$    (full red line) 
and
$\delta_{II}$ (dashed  blue line)
as function of the redshift, parameters as 
in Table  \ref{chi2darkunion}.
}
\label{percentageerror}
\end{figure}

We evaluate the integrand   
of the  integral (\ref{integralez}) 
with 
a first series expansion, $T_I$,  about $z=0$, denoted by $I$
and 
a second  series expansion,
 $T_{I}I$,
 about $z=1$, denoted by $II$.
The order of 
expansion  for the two series   is 7.
The  integration of  $T_I$ in z is denoted 
by $Iz_{I,7}$ and  gives  
\begin{equation}
Iz_{I,7}(z;\om,w)= \sum_{i=1}^{i=7} c_{I,i} z^i
\quad \,
\end{equation}
and the coefficients, $c_{I,i} $, are reported 
in  Appendix \ref{appendixa}.
The integral, $Iz_{II,7}$  of the 
second Taylor expansion  about $z=1$, $T_{II} $ 
is complicated and we limit ourselves to order 2, $Iz_{II,2}$,
see  Appendix  \ref{appendixa}.
The two definite integrals, 
 $F_{I,7}(z;\om,w)$ and   $F_{II,7}(z;\om,w)$
are 
\begin{equation}
F_{I,7}(z;\om,w)  = Iz_{I,7}(z=z;\om,w)  - Iz_{I,7}(z=0;\om,w)
\quad ,
\label{definitefz71}
\end{equation} 
and
\begin{equation}
F_{II,7}(z;\om,w)  = Iz_{II,7}(z=z;\om,w)  - Iz_{II,7}(z=0;\om,w)
\quad .
\label{definitefz72}
\end{equation} 
The percent error, $\delta$,
between the analytical integral $F$ and the two  approximations,
$F_{I,7}$ and   $F_{II,7}$   is evaluated  as
\begin{eqnarray}
\delta_I    = \Big | 1- \frac{F_{I,7}}{F} \Big | \times 100    \\
\delta_{II} = \Big | 1- \frac{F_{II,7}}{F}\Big | \times 100   
\quad .  
\end{eqnarray}
On inserting the astrophysical parameters
as reported in Table \ref{chi2darkunion}
we have $\delta_I=\delta_{II}$ at 
 $z \approx 0.58$,
see Figure \ref{percentageerror}. 

The above value in z will therefore be the boundary 
between   region I and 
region II for  the Taylor approximation
of the definite integral
\begin{equation}
F_{7}(z;\om,w)=
\left \{ \begin{array}{ll} 
F_{II,7}(z;\om,w), & 0.58 \leq z \leq 1.4    \\
F_{I,7}(z;\om,w) , & 0 < z < 0.58
\end {array}
\right \}
\label{definitefztaylor}
\end{equation}

\section{Variable equation of state}

\label{wzcdm}
The dark energy  as function of the redshift 
is  assumed to be 
\begin{equation}
w(z) = w_0 +w_1  \frac{z}{1+z}
\quad ,
\end{equation}
where $w_0$ and $w_1$ are two  parameters  to be fixed 
by the fit.
The  Hubble radius  in $wzCDM$ cosmology is  
\begin{eqnarray}
d_H(z;\om,w_0,w_1) =  \nonumber  \\
\frac {1}
{
\sqrt { \left( 1+z \right) ^{3}{\it \om}+ \left( 1-{\it \om} \right) 
 \left( 1+z \right) ^{3\,w_{{0}}+3\,w_{{1}}+3}{{\rm e}^{-3\,{\frac {w_
{{1}}z}{1+z}}}}}
}
\label{radiusvariable}
\end{eqnarray}
which is the same as  
equation(20) in \cite{Wei2015}.
The  above  integral does not yet have an analytical expression
and we  evaluate the integrand   
with 
a first series expansion about $z=0$ 
and 
a second  series expansion  about $z=1$.
Also here the order of the two series expansion is 7.
The  integration in $z$ is denoted 
by $Iwz_{I,7}$ and  gives  
\begin{equation}
Iwz_{I,7}(z;\om,w_0,w_1)= \sum_{i=1}^{i=7} c_{I,i} z^i
\quad \,
\label{iwztaylor}
\end{equation}
and the first five coefficients, $c_{I,i} $, are reported 
in  Appendix \ref{appendixc}.
The integral, $Iwz_{II,7}$  of the 
second Taylor expansion  about $z=1$  
is complicated and we limit ourselves to order 2, $Iwz_{II,2}$,
see  Appendix  \ref{appendixc}.
The two definite integrals, 
 $Fwz_{I,7}(z;\om,w_0,w_1)$ and   $Fwz_{II,7}(z;\om,w_0,w_1)$
are 
\begin{eqnarray}
Fwz_{I,7}(z;\om,w_0,w_1)  =  \nonumber \\
 Iwz_{I,7}(z=z;\om,w_0,w_1)  
- Iwz_{I,7}(z=0;\om,w_0,w_1)
\quad ,
\label{definitefwz71}
\end{eqnarray} 
and
\begin{eqnarray}
Fwz_{II,7}(z;\om,w_0,w_1)  =
\nonumber  \\
 Iwz_{II,7}(z=z;\om,w_0,w_1)  - 
Iwz_{II,7}(z=0;\om,w_0,w_1)
\quad .
\label{definitefwz72}
\end{eqnarray} 
Finally the definite integral, $Fwz$, is 
\begin{eqnarray}
Fwz_{7}(z;\om,w_0,w_1)=  \nonumber \\
\left \{ 
\begin{array}{ll} 
Fwz_{II,7}(z;\om,w_0,w_1) , & 0.58 \leq z \leq 1.4    \\
Fwz_{I,7} (z;\om,w_0,w_1) , & 0 < z < 0.58
\end{array}
\right \}
\label{definitefwztaylor}
\end{eqnarray}
The above definite integral  can also be evaluated in a numerical way,
\newline
 $Fwz_{num}(z;\om,w_0,w_1)$. 

\section{Cardassian cosmology}

\label{seccardassian}
In flat Cardassian cosmology the Hubble radius 
is 
\begin{equation}
d_H(z;\om,w,n) =
{\frac {1}{\sqrt { \left( 1+z \right) ^{3}{\it \om}+ \left( 1-{\it 
\om} \right)  \left( 1+z \right) ^{3\,n}}}}
\quad ,
\end{equation} 
where n is a variable parameter, n=0 means $\Lambda$CDM cosmology,
see equation (17) in \cite{Wei2015}.
The indefinite integral  in the  variable  $z$ of the 
above  Hubble radius, $Iz$,  is 
\begin{equation}
Iz(z;\om,n) =
\int d_H(z;\om,n)  dz 
\quad .
\label{integralezcard}
\end{equation}
Also here 
in order to solve the indefinite integral we  
perform a change of variable
$1+z=t^{1/3}$
\begin{equation}
Iz(t;\om,n) =\frac{1}{3}
\int \!{\frac {1}{\sqrt {-{t}^{n}{\it \om}+{\it \om}\,t+{t}^{n}}{t}^{2/3
}}}\,{\rm d}t
\quad .
\label{integraletcard}
\end{equation}
The indefinite integral is
\begin{equation}
Iz(t;\om,n) =
\frac
{
-2\,
{\mbox{$_2$F$_1$}\Big(1/2,- \left( 6\,n-6 \right) ^{-1};\,{\frac
{6\,n-7}{6\,n-6}};\,{\frac {{t}^{n-1} \left( {\it \om}-1 \right) }{{\it
\om}}}\Big)}
}
{
\sqrt {{\it \om}}\sqrt [6]{t}
}
\quad ,
\label{hubbleintegralcard}
\end{equation}
where ${\2F1(a,b;\,c;\,z)}$ is the
regularized hypergeometric
function.
We now return to the original variable $z$ 
as function of $z$ which is 
\begin{eqnarray}
Iz(z;\om,n) = \nonumber \\
\frac
{
-2\,
{\mbox{$_2$F$_1$}\Big(1/2,- \left( 6\,n-6 \right) ^{-1};\,{\frac
{6\,n-7}{6\,n-6}};
\,{\frac { \left(  \left( 1+z \right) ^{3} \right)^{n-1} 
\left( {\it \om} -1 \right) }{{\it \om}}}\Big)}
}
{
\sqrt {{\it \om}}\sqrt [6]{ \left( 1+z \right) ^{3}}
}
\quad .
\label{icardz}
\end{eqnarray}
We denote by $F_c(z;\om,n)$ the definite integral
\begin{equation}
F_c(z;\om,n)  = Iz(z=z;\om,n)  - Iz(z=0;\om,n)
\quad .
\label{definitefzcard}
\end{equation}

\section{The distance modulus}

\label{dm}
The luminosity distance, $\dl$,
for wCDM cosmology in the case of the
analytical solution  is  
\begin{equation}
  \dl(z;c,H_0,\om,w) = \frac{c}{H_0} (1+z) F(z;\om,w)
\quad ,
\end{equation}
where $F(z;\om,w)$  is  given by  
equation (\ref{definitefz}) 
and  in the case of the Taylor  approximation is
\begin{equation}
  \dlsette  (z;c,H_0,\om,w) = \frac{c}{H_0} (1+z) F_{7}(z;\om,w)
\quad ,
\end{equation}
where $F_7(z;\om,w)$  is  given by  
equation (\ref{definitefztaylor}). 
The distance modulus 
in the case of the analytical solution for wCDM is 
\begin{equation}
(m-M) =25 +5 \log_{10}\bigg ( \dl(z;c,H_0,\om,w)     \bigg)
\quad ,
\label{distmodhyper}
\end{equation}
and in the case of the Taylor approximation
\begin{equation}
(m-M)_7 =25 +5 \log_{10}\bigg ( \dlsette(z;c,H_0,\om,w)     \bigg)
\quad .
\label{distmodtaylor}
\end{equation}
In the case of variable equation of state, $wzCDM$,
the numerical luminosity distance is 
\begin{equation}
  \dlnum  (z;c,H_0,\om,w_0,w_1) = \frac{c}{H_0} (1+z)
Fwz_{num}(z;\om,w_0,w_1)
\quad ,
\end{equation}
where $Fwz_{num}(z;\om,w_0,w_1)$ is the definite numerical integral
and the Taylor  approximation for the luminosity distance  is  
\begin{equation}
  \dlsette  (z;c,H_0,\om,w_0,w_1) = \frac{c}{H_0} (1+z)
Fwz_{7}(z;\om,w_0,w_1)
\quad ,
\end{equation}
where $Fwz_7(z;\om,w_0,w_1)$  is  given by  
equation (\ref{definitefwztaylor}).    
In $wzCDM$ the numerical distance modulus  is
\begin{equation}
(m-M)_{num} =25 +5 \log_{10}\bigg ( \dlnum(z;c,H_0,\om,w_0,w_1)     \bigg)
\quad ,
\label{distmonumericalwz}
\end{equation}
and the Taylor approximated distance modulus is 
\begin{equation}
(m-M)_7 =25 +5 \log_{10}\bigg ( \dlsette(z;c,H_0,\om,w_0,w_1)     \bigg)
\quad .
\label{distmodnumericalwz}
\end{equation}
In the case of Cardassian cosmology
the luminosity distance is 
\begin{equation}
  \dl(z;c,H_0,\om,n) = \frac{c}{H_0} (1+z) F_c(z;\om,n)
\quad ,
\end{equation}
where $F_c(z;\om,n)$  is  given by  
equation (\ref{definitefzcard})
and the distance modulus is
\begin{equation}
(m-M) =25 +5 \log_{10}\bigg ( \dl(z;c,H_0,\om,n)     \bigg)
\quad .
\label{distmodhyperwz}
\end{equation}

\begin{figure}
\includegraphics[width=10cm,angle=-90]{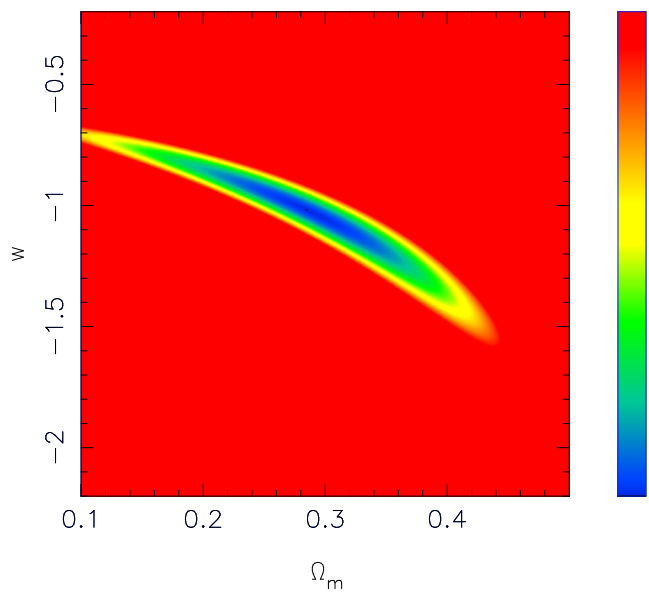}
\caption{
Map of the $\chi^2$ 
in wCDM  cosmology
when  $H_0=(70.02\pm 0.35)$. 
}
\label{chi2_dark}
\end{figure}

\begin{figure}
\includegraphics[width=10cm,angle=-90]{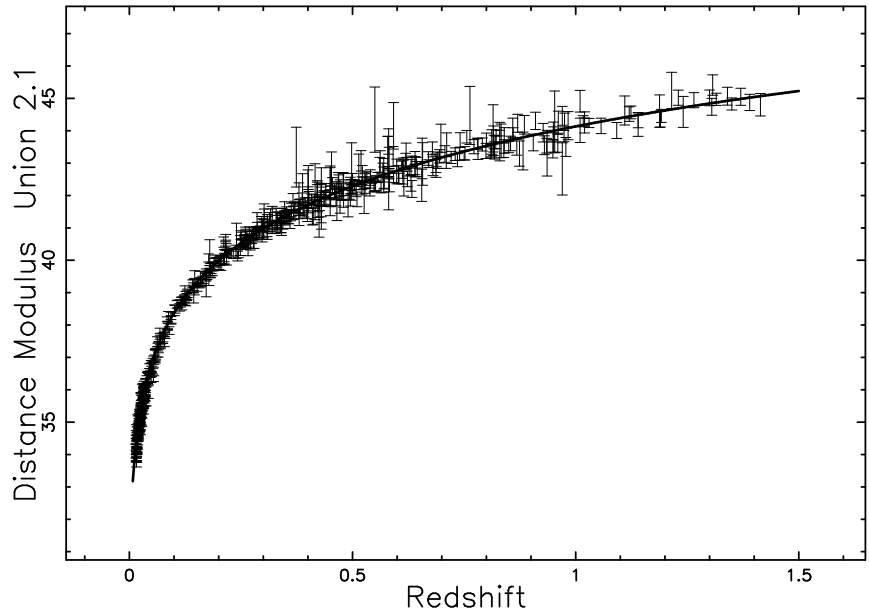}
\caption{
Hubble diagram for the  Union 2.1  compilation.
The solid line represents the best fit
for the exact distance modulus  in wCDM  cosmology 
as represented by Eq.~(\ref{distmodhyper}).
Parameters as in third  line of Table 
\ref{chi2darkunion}; Union 2.1  compilation.
}
\label{distmdarkunion21}
\end{figure}

The  cosmological parameters unknown are
three, $H_0,\om$ and $w$, 
in the case of wCDM  and
four,  
$H_0,\om$ ,  $w_0$ and $w_1$, 
in the case of $wzCDM$.
In flat Cardassian cosmology 
the number of parameters is three, $H_0,\om$ and $n$.
In presence of a given sample for the distance
modulus we can map the chi-square 
as  given by formula (\ref{chiquaredef}),
see Figure \ref{chi2_dark} in the case 
of wCDM with hypergeometric solution.
The above cosmological parameters  
are obtained by a fit of  the astronomical 
data  for the distance modulus of  SNs
via  
the
Levenberg--Marquardt  method
(subroutine MRQMIN in \cite{press})
which  minimizes the 
chi-square 
as  given by formula (\ref{chiquaredef}).
Table \ref{chi2darkunion} presents the above cosmological 
parameters 
for  the  Union 2.1  compilation of SNs  and
Figure \ref{distmdarkunion21} 
reports  the best  fit.
As a  practical example  of the utility of the 
cosmological parameters determination 
we report the distance modulus in an explicit form
for  the Union 2.1  compilation 
in wCDM 
\begin{eqnarray} 
(m-M)=
5+5\,{\frac {1}{\ln    ( 10  ) }} \times
\nonumber \\
{
\ln \Big   (  4281.52\, 
   ( 1+z  )    
} \times
\nonumber \\
{
( - 3.8\,{\frac {
{\mbox{$_2$F$_1$}\big( 0.1661,\frac{1}{2};\, 1.1661;\,- 2.6101\,   (
{z}^{3}+3\,{z}^{2}+3\,z+1  ) ^{- 1.003}\big)}
}{\sqrt [6]{{z}^{3}+3\,{z}^{2}+3\,z+1}}}
}
\times
\nonumber  \\
{
+ 3.4146  ) 
  \Big ) }  \\
when ~ 0 < z<1.4   
\quad  ,
\nonumber 
\end{eqnarray}
and in  flat Cardassian cosmology
\begin{eqnarray} 
(m-M)= \frac{1}{\ln    ( 10   ) }
25\,\ln    ( 10   ) 
\nonumber \\ \times
+5\,\ln  \Bigg  ( - 4273.59\, \Big  ( 1+z
   )    (  3.62142\,   ( {z}^{3}+3\,{z}^{2}+3\,z+1
   ) ^{- 0.16666}
\times
\nonumber \\
{\mbox{$_2$F$_1$}\big ( 0.15417,1/2;\, 1.1541;\,- 2.2786\,   (
{z}^{3}+3\,{z}^{2}+3\,z+1   ) ^{- 1.081}\big)}
- 3.304  \Big ) \Bigg   ) 
  \\
when ~ 0 < z<1.4   
\quad  .
\nonumber 
\end{eqnarray}

\begin{table}[ht!]
\caption
{
Numerical values 
from  the  Union 2.1  compilation
of
$\chi^2$,
$\chi_{red}^2$
and 
$Q$, where 
$k$ stands for the number of parameters.
}
\label{chi2darkunion}
\begin{center}
\resizebox{12cm}{!}
{
\begin{tabular}{|c|c|c|c|c|c|c|}
\hline
Cosmology  &  SNs& $k$    &   parameters    & $\chi2$& $\chi_{red}^2$
&
$Q$      \\
\hline
$\Lambda$CDM & 580 &  3
& $H_0$ = 69.81; $\om=0.239$; $\ola=0.651$
& 562.61 &  0.975  & 0.658 \\
wCDM Hypergeometric solution 
& 580  
& 3
& $H_0=(70.02\pm 0.35)$; 
$\om =(0.277\pm 0.025)$ 
; $w=  (-1.003\pm 0.05)$ 
&  562.21 &  0.974 & 0.662 
\\
wCDM Taylor approximation  
& 580  
& 3
& $H_0=(70.02\pm 0.47)$; 
$\om =(0.282 \pm 0.07)$ 
; $w=  (-1.01\pm 0.2)$ 
&  562.21 &  0.974 & 0.662 
\\
$wzCDM$ Taylor approximation  
& 580  
& 4
& $H_0=(70.08\pm 0.31 )$; 
$\om =(0.284 \pm 0.01)$; 
$w_0=  (-1.03\pm 0.031)$;
$w_1=  ( 0.1 \pm 0.018)$; 
&  562.21 &  0.976 & 0.651 
\\
Cardassian 
& 58k0  
& 3
& $H_0=(70.15  \pm 0.38 )$; 
$\om =(0.305   \pm 0.019)$ 
; $n=  (-0.081 \pm 0.01 )$ 
&  562.35 &  0.974 & 0.661 
\\
\hline

\end{tabular}
}
\end{center}
\end{table}

\begin{figure}
\includegraphics[width=10cm,angle=-90]{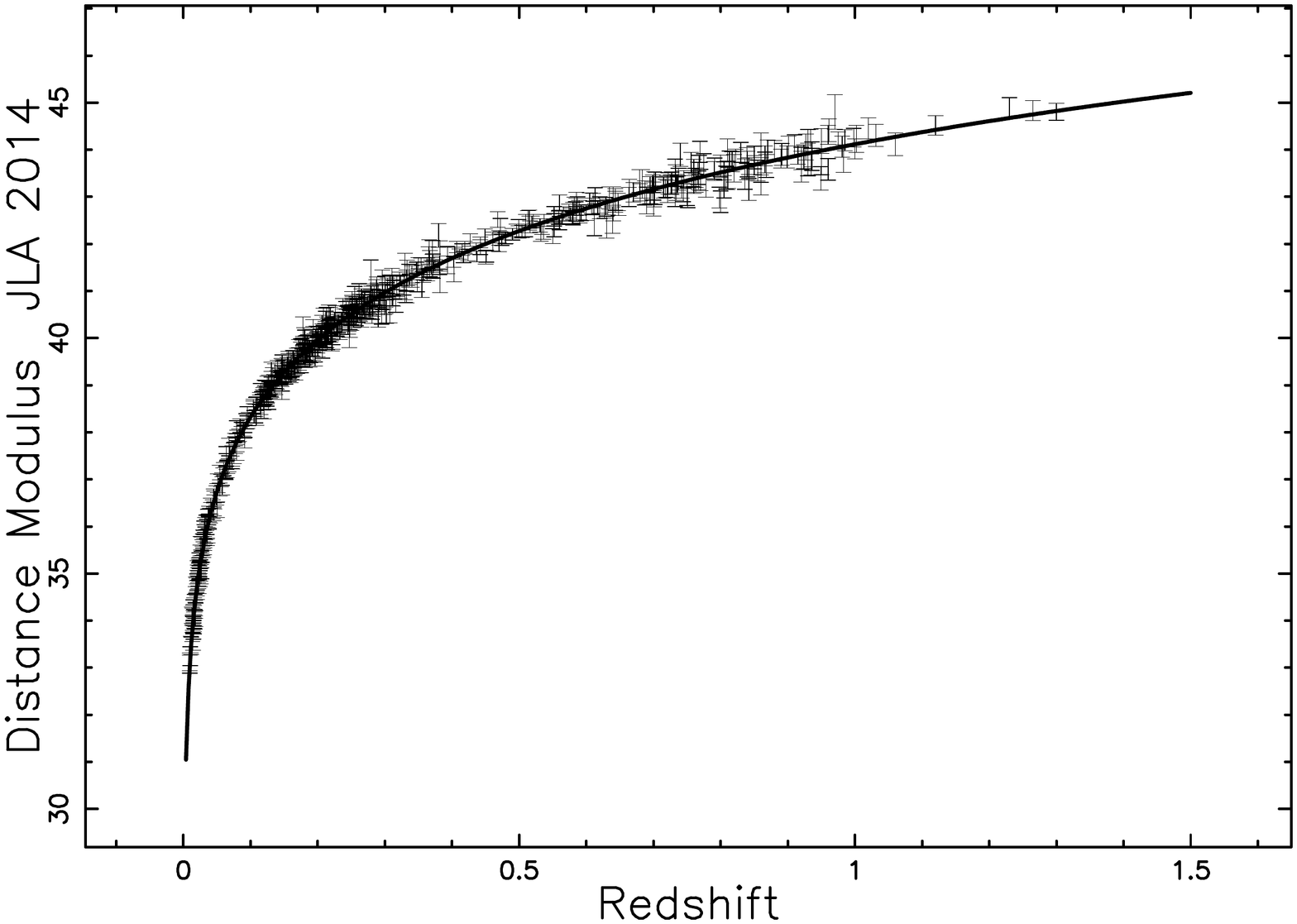}
\caption{
Hubble diagram for the  JLA  compilation.
The solid line represents the best fit
for the exact distance modulus  in wCDM  cosmology 
as represented by equation~(\ref{distmodhyper}).
Parameters as in the third  line of Table 
\ref{chi2darkjla}; JLA   compilation.
}
\label{distmdarkjla}
\end{figure}

Table  \ref{chi2darkjla} reports the cosmological 
parameters for  the JLA compilation 
and Figure \ref{distmdarkjla}
the connected fit.
\begin{table}[ht!]
\caption
{
Numerical values 
for  the  JLA  compilation
of
$\chi^2$,
$\chi_{red}^2$
and 
$Q$, where 
$k$ stands for the number of parameters.
}
\label{chi2darkjla}
\begin{center}
\resizebox{12cm}{!}
{
\begin{tabular}{|c|c|c|c|c|c|c|}
\hline
Cosmology  &  SNs& $k$    &   parameters    & $\chi2$& $\chi_{red}^2$
&
$Q$      \\
\hline
$\Lambda$CDM & 740 &  3
& $H_0$ = 69.39; $\om=0.18$; $\ola=0.537$
& 625.74 &  0.849  & 0.99 \\
wCDM Hypergeometric solution 
& 740
&  3
& $H_0=(69.71\pm 0.5  )$; 
$\om =(0.293 \pm 0.021)$ 
; $w=  (-0.996\pm 0.08)$ 
&  627.908 &  0.851 & 0.998 
\\
wCDM Taylor approximation 
& 740
&  4
& $H_0=(69.99\pm 0.29  )$; 
$\om =(0.133 \pm 0.13)$ 
; $w=  (-0.709\pm 0.18)$ 
&  625.69 &  0.848 & 0.998 
\\
$wzCDM$ Taylor approximation  
& 740  
& 4
& $H_0=(69.99 \pm 0.29 )$; 
$\om =(0.3  \pm 0.009)$; 
$w_0=  (-1.05\pm 0.027)$;
$w_1=  ( 0.097 \pm 0.01)$; 
&  628.76 &  0.854 & 0.998 
\\
Cardassian 
& 740  
& 3
& $H_0=(70.036  \pm 0.44 )$; 
$\om =(0.301    \pm 0.019)$; 
$n=  (-0.055  \pm 0.0045 )$ 
&  628.73 &  0.863 & 0.999 
\\
\hline
\end{tabular}
}
\end{center}
\end{table}

\begin{figure}
\includegraphics[width=10cm,angle=-90]{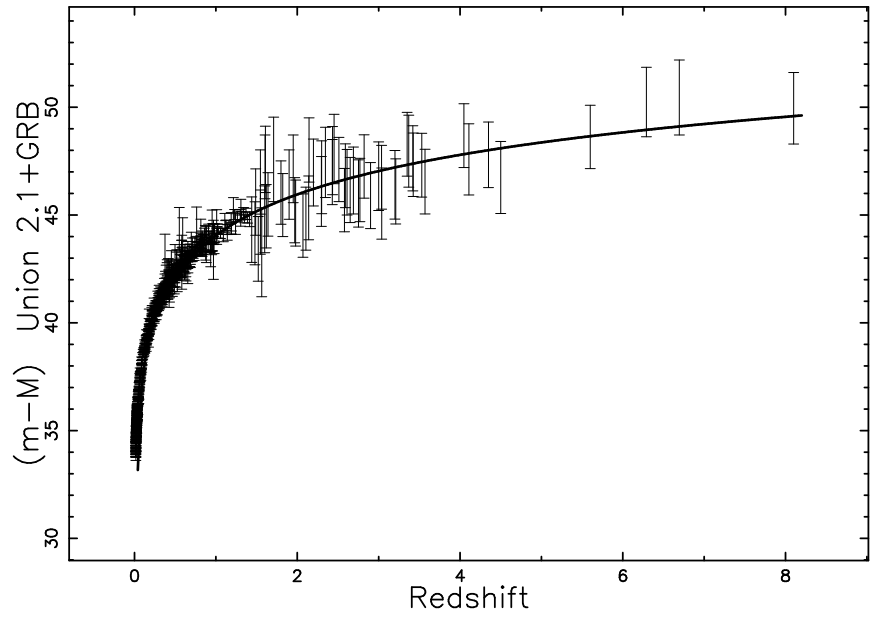}
\caption{
Hubble diagram for the Union 2.1  compilation + 
the "Hymnium" GRBs sample.
The solid line represents the best fit
for the exact distance modulus  in wCDM  cosmology 
as represented by Eq.~(\ref{distmodhyper}).
Parameters as in second  line of Table 
\ref{chi2darkgrb}.
}
\label{distmdarkgrb}
\end{figure}

The presence  of  the
"Hymnium" GRBs sample  allows to calibrate the distance
modulus in the high redshift region, see 
Table \ref{chi2darkgrb}  and 
Figure \ref{distmdarkgrb}.
\begin{table}[ht!]
\caption
{
Numerical values 
from  the  Union 2.1  compilation + 
the "Hymnium"  GRBs sample 
of
$\chi^2$,
$\chi_{red}^2$
and 
$Q$, where 
$k$ stands for the number of parameters.
}
\label{chi2darkgrb}
\begin{center}
\resizebox{12cm}{!}
{
\begin{tabular}{|c|c|c|c|c|c|c|}
\hline
Cosmology  &  SNs& $k$    &   parameters    & $\chi2$& $\chi_{red}^2$
&
$Q$      \\
\hline
$\Lambda$CDM & 639 &  3
& $H_0$ = 69.80; $\om=0.239$; $\ola=0.651$
& 586.08 &  0.921  & 0.922 \\
wCDM Hypergeometric solution 
& 639
& 3
& $H_0=(70.12\pm 0.4)$; 
$\om =(0.294\pm 0.024)$ 
; 
$w=  (-1.04\pm 0.04)$ 
&  585.42 & 0.92 &  0.924 
\\
$wzCDM$ numerical integration  
& 639  
& 4
& $H_0=(70 \pm 0.32 )$; 
$\om =(0.3  \pm 0.011)$; 
$w_0=  (-1.05\pm 0.033)$;
$w_1=  ( 0.1 \pm 0.01)$; 
&  585.59 &  0.922 & 0.92 
\\
Cardassian 
& 639  
& 3
& $H_0=(70.10  \pm 0.42 )$; 
$\om =(0.299   \pm 0.019)$ 
; $n=  (-0.063 \pm 0.0095 )$ 
&  585.43 &  0.92 & 0.924 
\\
\hline
\end{tabular}
}
\end{center}
\end{table}

The extension of the Hubble diagram to the GRBs, as an example, has been
implemented 
in  \cite{Liang2008,Wei2010,Lin2016,Gupta2019a,Marosi2019}.

\section{Conclusions}

{\bf Constant equation of state}

In the case of wCDM cosmology we found a new  analytical expression
for the Hubble distance in terms of the hypergeometric function,
see  equation (\ref{hubbleintegral}).
As a consequence an analytical expression for the luminosity distance
and the distance modulus is derived.
Two approximate Taylor expansions for  the Hubble distance  about 
$z=0$ and $z=1$ of order 7 are also derived.
The derivation  of the value of $w$, $\om$  and $H_0$, here considered as
a parameter to be found, is given 
for the Union 2.1 compilation, the JLA compilation 
and the Union 2.1 compilation plus the "Hymnium" GRBs sample,
see Tables 
\ref{chi2darkunion}, 
\ref{chi2darkjla}  
and \ref{chi2darkgrb}.  
As an example, in the case of the Union 2.1 compilation,
we have derived 
$H_0=(70.02\pm 0.35)$, $\om =(0.277\pm 0.025)$ and 
$w=  (-1.003\pm 0.05)$.

{\bf Variable equation of state}
  
In the case of $wzCDM$ cosmology  the Hubble distance,
equation (\ref{radiusvariable}) is evaluated numerically
and with a Taylor expansion of order 7,
see equation (\ref{iwztaylor}).
The four parameters $w_0$, $w_1$, $\om$  and $H_0$
are reported 
in  Tables \ref{chi2darkunion}, \ref{chi2darkjla}  
and \ref{chi2darkgrb}.
As an example, in the case of the Union 2.1 compilation,
we have found 
 $H_0=(70.08\pm 0.31 )$, 
$\om =(0.284 \pm 0.01)$, 
$w_0=  (-1.03\pm 0.031)$,
and 
$w_1=  ( 0.1 \pm 0.018)$.

{\bf High redshift}
The inclusion   of  the
"Hymnium" GRBs sample  allows 
to extend  the calibration of
the distance modulus up to $z$=8, see 
Table \ref{chi2darkgrb}.
As an example the Union 2.1  compilation + 
the "Hymnium"  GRBs sample gives  
$H_0=(70 \pm 0.32 )$, 
$\om =(0.3  \pm 0.011)$, 
$w_0=  (-1.05\pm 0.033)$,
and
$w_1=  ( 0.1 \pm 0.01)$. 

{\bf Cardassian cosmology}

A new solution for the Hubble radius for Cardassian cosmology 
is presented in terms
of the hypergeometric function, see equation (ref{icardz}).
As an example, in the case of the Union 2.1 compilation,
we have derived 
$H_0=(70.15  \pm 0.38 )$, 
$\om =(0.305   \pm 0.019)$ and 
$n=  (-0.081 \pm 0.01 )$.

\appendix
\setcounter{equation}{0}

\renewcommand{\theequation}{\thesection.\arabic{equation}}

\section{Taylor expansion when w is constant} 
\label{appendixa}
The coefficients of the Taylor expansion 
of $Iz_{I,7}(z;\om,w)$ about $z=0$ 
\begin{equation}
c_{I,1}=
1
\quad ,
\end{equation}
\begin{equation}
c_{I,2}=
3/4\,w{\it \om}-3/4\,w-3/4
\quad ,
\end{equation}
\begin{eqnarray}
c_{I,3}=
-3/2\,{\it \om}\,{w}^{2}-w{\it \om}+3/8\,{w}^{2}+w+5/8+
\nonumber \\
{\frac {9\,{{\it 
\om}}^{2}{w}^{2}}{8}}
\quad ,
\end{eqnarray}

\begin{eqnarray}
c_{I,4}=
-{\frac {71\,w}{64}}-{\frac {9\,{w}^{3}}{64}}-{\frac{35}{64}}-{\frac {
45\,{w}^{2}}{64}}+{\frac {45\,{\it \om}\,{w}^{2}}{16}}-{\frac {135\,{{
\it \om}}^{2}{w}^{2}}{64}}
\nonumber \\
-{\frac {243\,{{\it \om}}^{2}{w}^{3}}{64}}+{
\frac {117\,{\it \om}\,{w}^{3}}{64}}+{\frac {135\,{{\it \om}}^{3}{w}^{3}
}{64}}+{\frac {71\,w{\it \om}}{64}}
\quad ,
\end{eqnarray}

\begin{eqnarray}
c_{I,5}=
{\frac {93\,w}{80}}+{\frac{63}{128}}+{\frac {27\,{w}^{3}}{80}}+{\frac 
{27\,{w}^{4}}{640}}+{\frac {309\,{w}^{2}}{320}}-{\frac {309\,{\it \om}
\,{w}^{2}}{80}}
\nonumber \\
+{\frac {927\,{{\it \om}}^{2}{w}^{2}}{320}}+{\frac {729
\,{{\it \om}}^{2}{w}^{3}}{80}}-{\frac {351\,{\it \om}\,{w}^{3}}{80}}-{
\frac {81\,{{\it \om}}^{3}{w}^{3}}{16}}
\nonumber \\
+{\frac {2349\,{{\it \om}}^{2}{w}
^{4}}{320}}-{\frac {27\,{\it \om}\,{w}^{4}}{16}}-{\frac {81\,{{\it \om}}
^{3}{w}^{4}}{8}}+{\frac {567\,{{\it \om}}^{4}{w}^{4}}{128}}
\nonumber \\
-{\frac {93
\,w{\it \om}}{80}}
\quad ,
\end{eqnarray}

\begin{eqnarray}
c_{I,6}=
-{\frac {3043\,w}{2560}}-{\frac{231}{512}}-{\frac {27\,{w}^{5}}{2560}}
-{\frac {141\,{w}^{3}}{256}}-{\frac {63\,{w}^{4}}{512}}
\nonumber \\
-{\frac {14175
\,{{\it \om}}^{4}{w}^{5}}{512}}
+{\frac {5103\,{{\it \om}}^{5}{w}^{5}}{
512}}-{\frac {301\,{w}^{2}}{256}}+{\frac {301\,{\it \om}\,{w}^{2}}{64}}
\nonumber \\
-{\frac {903\,{{\it \om}}^{2}{w}^{2}}{256}}-{\frac {3807\,{{\it \om}}^{2
}{w}^{3}}{256}}
+{\frac {1833\,{\it \om}\,{w}^{3}}{256}}+{\frac {2115\,{
{\it \om}}^{3}{w}^{3}}{256}}
\nonumber \\
-{\frac {5481\,{{\it \om}}^{2}{w}^{4}}{256}}
+{\frac {315\,{\it \om}\,{w}^{4}}{64}}+{\frac {945\,{{\it \om}}^{3}{w}^{
4}}{32}}-{\frac {6615\,{{\it \om}}^{4}{w}^{4}}{512}}
\nonumber \\
-{\frac {2673\,{{
\it \om}}^{2}{w}^{5}}{256}}+{\frac {3267\,{\it \om}\,{w}^{5}}{2560}}+{
\frac {6885\,{{\it \om}}^{3}{w}^{5}}{256}}+{\frac {3043\,w{\it \om}}{
2560}}
\quad ,
\end{eqnarray}

\begin{eqnarray}
c_{I,7}=
{\frac {2689\,w}{2240}}+{\frac {81\,{w}^{6}}{35840}}+{\frac {81\,{w}^{
5}}{2240}}+{\frac {171\,{w}^{3}}{224}}+{\frac {1665\,{w}^{4}}{7168}}
\nonumber \\
+{
\frac {48259\,{w}^{2}}{35840}}+{\frac{429}{1024}}+{\frac {95985\,{{
\it \om}}^{4}{w}^{6}}{1024}}-{\frac {19683\,{{\it \om}}^{5}{w}^{6}}{256}
}
\nonumber \\
+{\frac {24057\,{{\it \om}}^{6}{w}^{6}}{1024}}+{\frac {61479\,{{\it \om
}}^{2}{w}^{6}}{5120}}-{\frac {1053\,{\it \om}\,{w}^{6}}{1280}}
\nonumber  \\
-{\frac {
23085\,{{\it \om}}^{3}{w}^{6}}{448}}+{\frac {6075\,{{\it \om}}^{4}{w}^{5
}}{64}}-{\frac {2187\,{{\it \om}}^{5}{w}^{5}}{64}}+{\frac {8019\,{{\it 
\om}}^{2}{w}^{5}}{224}}
\nonumber \\
-{\frac {9801\,{\it \om}\,{w}^{5}}{2240}}-{\frac 
{20655\,{{\it \om}}^{3}{w}^{5}}{224}}+{\frac {144855\,{{\it \om}}^{2}{w}
^{4}}{3584}}-{\frac {8325\,{\it \om}\,{w}^{4}}{896}}
\nonumber  \\
-{\frac {24975\,{{
\it \om}}^{3}{w}^{4}}{448}}+{\frac {24975\,{{\it \om}}^{4}{w}^{4}}{1024}
}+{\frac {4617\,{{\it \om}}^{2}{w}^{3}}{224}}-{\frac {2223\,{\it \om}\,{
w}^{3}}{224}}
\nonumber  \\
-{\frac {2565\,{{\it \om}}^{3}{w}^{3}}{224}}-{\frac {48259
\,{\it \om}\,{w}^{2}}{8960}}+{\frac {144777\,{{\it \om}}^{2}{w}^{2}}{
35840}}-{\frac {2689\,w{\it \om}}{2240}}
\quad .
\end{eqnarray}

The integral of the Taylor expansion
of order 2 about  $z=1$  is 
\begin{equation}
Iz_{II,2} = \frac{N}{D}
\quad ,
\end{equation}
where  
\begin{eqnarray}
N=
\Big ( 3\,{8}^{w}{\it \om}\,wz-6\,{8}^{w}w{\it \om}+3\,{8}^{w}{\it \om}
\,z-3\,wz{8}^{w}
\nonumber  \\
-14\,{8}^{w}{\it \om}+6\,w{8}^{w}-3\,z{8}^{w}-3\,{\it 
\om}\,z+14\,{8}^{w}+14\,{\it \om} \Big) z
\end{eqnarray}
and  
\begin{equation}
D=
\left( -{2}^{3+3\,w}{\it \om}+{2}^{3+3\,w}+8\,{\it \om} \right) ^{3/2}
\quad .
\end{equation}

\setcounter{equation}{0}
\section{The hypergeometric function}

\label{appendixb}

The   regularized hypergeometric
function,
 ${\2F1(a,b;\,c;\,z)}$, 
as defined by the Gauss series, is  
\begin{eqnarray}
\2F1(a,b;\,c;\,z)=
\sum_{s=0}^{\infty}\frac{{\left(a\right)_{s}}
{\left(b\right)_{s}}}{{\left(c\right)_{s}}s!}z^{s}
=1+\frac{ab}{c}z+\frac{a(a+1)b(b+1)}{c(c+1)2!}z^{2}+\cdots
\nonumber \\
=\frac{\Gamma\left(c\right)}{\Gamma\left(a\right)\Gamma
\left(b\right)}\sum_{s=0}^{\infty}\frac{\Gamma\left(a+s\right)\Gamma\left(b+s
\right)}{\Gamma\left(c+s\right)s!}z^{s}
\end{eqnarray} 
where
$z=x+iy$, 
$(a)_s$
is the Pochhammer symbol
\begin{equation}
(a)_s=a(a+1)\dots(a+s-1)
\quad  ,
\end{equation}
 $\Gamma\left(z\right)$ is the Gamma function defined as  
\begin{equation}
\Gamma\left(z\right)=\int_{0}^{\infty}e^{-t}t^{z-1}\mathrm{d}t
\quad ,
\end{equation}
$z$ is  a complex  variable defined on the disk 
$|z| <1$  that should not be confused with the redshift, see  
\cite{Abramowitz1965,Seggern1992,Thompson1997,Gradshteyn2007,NIST2010}.
The following  relationship
\begin{equation}
{\mbox{$_2$F$_1$}(a,b;\,c;\,x)}
=
\left( 1-x \right) ^{-a}
{\mbox{$_2$F$_1$}(a,c-b;\,c;\,{\frac {x}{x-1}})}
\end{equation}
connect the the hypergeometric function 
with x in (-1,1) to one 
with x in ($-\infty,\frac{1}{2})$,
see more  details in \cite{Oldham2010}.

\setcounter{equation}{0}
\section{Taylor expansion when w is variable} 
\label{appendixc}
The coefficients of the Taylor expansion 
of $Iwz_{I,7}(z;\om,w_0,w_1)$ about $z=0$ 
\begin{equation}
c_{I,1}=
1
\quad ,
\end{equation}

\begin{equation}
c_{I,2}=
\frac{3}{4}\,w_{{0}}{\it \om}-\frac{3}{4}\,w_{{0}}-\frac{3}{4}
\quad ,
\end{equation}

\begin{eqnarray}
c_{I,3}=
5/8+w_{{0}}-1/4\,w_{{1}}+1/4\,w_{{1}}{\it \om}-w_{{0}}{\it \om}+3/8\,{w_
{{0}}}^{2}-3/2\,{\it \om}\,{w_{{0}}}^{2}
\nonumber  \\
+{\frac {9\,{{\it \om}}^{2}{w_{{0
}}}^{2}}{8}}
\quad ,
\end{eqnarray}

\begin{eqnarray}
c_{I,4}=
-{\frac{35}{64}}-{\frac {71\,w_{{0}}}{64}}+{\frac {17\,w_{{1}}}{32}}-{
\frac {17\,w_{{1}}{\it \om}}{32}}+{\frac {71\,w_{{0}}{\it \om}}{64}}-{
\frac {45\,{w_{{0}}}^{2}}{64}}
\nonumber \\
+{\frac {9\,w_{{0}}w_{{1}}}{32}}+
{\frac 
{45\,{\it \om}\,{w_{{0}}}^{2}}{16}}
-{\frac {135\,{{\it \om}}^{2}{w_{{0}}
}^{2}}{64}}
-{\frac {243\,{{\it \om}}^{2}{w_{{0}}}^{3}}{64}}+{\frac {117
\,{\it \om}\,{w_{{0}}}^{3}}{64}}+
\nonumber \\
{\frac {135\,{{\it \om}}^{3}{w_{{0}}}^{
3}}{64}}-
{\frac {9\,{w_{{0}}}^{3}}{64}}-
{\frac {9\,{\it \om}\,w_{{0}}w_
{{1}}}{8}}
+{\frac {27\,{{\it \om}}^{2}w_{{0}}w_{{1}}}{32}}
\quad ,
\end{eqnarray}

\begin{eqnarray}
c_{I,5}=
{\frac {27\,{w_{{0}}}^{3}}{80}}+{\frac{63}{128}}-{\frac {9\,{w_{{1}}}^
{2}{\it \om}}{40}}+{\frac {2349\,{{\it \om}}^{2}{w_{{0}}}^{4}}{320}}-{
\frac {27\,{\it \om}\,{w_{{0}}}^{4}}{16}}
\nonumber \\
+{\frac {27\,{w_{{1}}}^{2}{{
\it \om}}^{2}}{160}}
-{\frac {81\,{{\it \om}}^{3}{w_{{0}}}^{4}}{8}}
+{
\frac {567\,{{\it \om}}^{4}{w_{{0}}}^{4}}{128}}-{\frac {27\,{w_{{0}}}^{
2}w_{{1}}}{160}}+{\frac {309\,{w_{{0}}}^{2}}{320}}
\nonumber\\
-\frac{3}{4}\,w_{{0}}w_{{1}}
+{\frac {729\,{{\it \om}}^{2}{w_{{0}}}^{3}}{80}}
-{\frac {351\,{\it \om}
\,{w_{{0}}}^{3}}{80}}
-{\frac {81\,{{\it \om}}^{3}{w_{{0}}}^{3}}{16}}
+{
\frac {93\,w_{{0}}}{80}}-{\frac {129\,w_{{1}}}{160}}
\nonumber \\
+{\frac {9\,{w_{{1
}}}^{2}}{160}}+{\frac {27\,{w_{{0}}}^{4}}{640}}
+{\frac {351\,{\it \om}
\,{w_{{0}}}^{2}w_{{1}}}{160}}
+{\frac {129\,w_{{1}}{\it \om}}{160}}
-{
\frac {309\,{\it \om}\,{w_{{0}}}^{2}}{80}}
\nonumber  \\
+{\frac {927\,{{\it \om}}^{2}{
w_{{0}}}^{2}}{320}}
-{\frac {93\,w_{{0}}{\it \om}}{80}}+{\frac {81\,{{
\it \om}}^{3}{w_{{0}}}^{2}w_{{1}}}{32}}
-{\frac {729\,{{\it \om}}^{2}{w_{
{0}}}^{2}w_{{1}}}{160}}
\nonumber \\
-\frac{9}{4}\,{{\it \om}}^{2}w_{{0}}w_{{1}}
+3\,{\it \om}
\,w_{{0}}w_{{1}}
\, .
\end{eqnarray}
The integral of the Taylor expansion
of order 2 about  $z=1$
in the case wzLCDM cosmology 
\begin{equation}
Iwz_{II,2} = \frac{Nwz}{Dwz}
\quad ,
\end{equation}
where  
\begin{eqnarray}
Nwz=
{{\rm e}^{\frac{3}{4}\,w_{{1}}}} \Big( 6\,{2}^{1/2+3\,w_{{0}}+3\,w_{{1}}}{
\it \om}\,zw_{{0}}+3\,{2}^{1/2+3\,w_{{0}}+3\,w_{{1}}}{\it \om}\,zw_{{1}}
\nonumber \\
-6\,{{\rm e}^{3/2\,w_{{1}}}}{\it \om}\,\sqrt {2}z
+6\,{2}^{1/2+3\,w_{{0}
}+3\,w_{{1}}}{\it \om}\,z-12\,{2}^{1/2+3\,w_{{0}}+3\,w_{{1}}}w_{{0}}{
\it \om}
\nonumber \\
-6\,{2}^{1/2+3\,w_{{0}}+3\,w_{{1}}}{\it \om}\,w_{{1}}
-6\,{2}^{1/
2+3\,w_{{0}}+3\,w_{{1}}}zw_{{0}}
-3\,{2}^{1/2+3\,w_{{0}}+3\,w_{{1}}}zw_
{{1}}
\nonumber \\
+28\,{{\rm e}^{3/2\,w_{{1}}}}{\it \om}\,\sqrt {2}
-28\,{2}^{1/2+3\,
w_{{0}}+3\,w_{{1}}}{\it \om}
\nonumber \\
-6\,{2}^{1/2+3\,w_{{0}}+3\,w_{{1}}}z
+12\,{2
}^{1/2+3\,w_{{0}}+3\,w_{{1}}}w_{{0}}
\nonumber \\
+6\,{2}^{1/2+3\,w_{{0}}+3\,w_{{1}}
}w_{{1}}+28\,{2}^{1/2+3\,w_{{0}}+3\,w_{{1}}} \Big ) z
\end{eqnarray}
and
\begin{eqnarray}
Dwz=
64\, \left( -{\it \om}\,{2}^{3\,w_{{0}}+3\,w_{{1}}}+{2}^{3\,w_{{0}}+3\,
w_{{1}}}+{\it \om}\,{{\rm e}^{3/2\,w_{{1}}}} \right) ^{3/2}
\quad  .
\end{eqnarray}

\providecommand{\newblock}{}

\providecommand{\newblock}{}
\end{document}